**Title page**

**Fluid laminarization in protein-based high internal phase emulsions process**

*Liang Guo[a]\*, Zi-an Deng[c], Yue-cheng Meng[b], Jing Chen[a], Sheng Fang,[b] Yang Pan[a] and Jie Chen[b]*

a   School of pharmacy, Nanjing University of Chinese medicine, 138 Xianlin Avenue, Nanjing, 210023, China

b   School of Food Science and Biotechnology, Zhejiang Gongshang University, 18 Xuezheng Road, Hangzhou, 310018, China

c   College of Agriculture & Biotechnology, Zhejiang University, 866 Yuhangtang Road, Hangzhou, 310058, China

\* Corresponding author

mingfangguo@njucm.edu.cn (L. Guo)

**Abstract**

Protein-based high internal phase emulsions (HIPEs) have gained tremendous attention in diverse fields, but their mechanism in the emulsification process remains elusive. In this article, HIPEs were stabilized directly by food-grade proteins, depending on a self-organized process featuring a fluid laminarization. We elucidated that the emulsification with the rotor-stator mixer is a typical non-equilibrium process. The crucial factor for the process is related to the irreversible energy dissipation, while the internal phase volume fraction is the threshold determining the laminarization. The feasible explanation speculated that the transition corresponds to the dissipative structure, i.e., compressive droplets, arising from the spatiotemporal self-organization, to dissipate the turbulent kinetic energy. We found a new paradigm of dissipative structure, comprehending such structure in the HIPEs emulsification process, which is expected to pave the way for its industrial-scale production with the virtue of low-cost proteins.

# 1 Introduction

High internal phase emulsions (HIPEs) are concentrated systems with a large volume of internal phase (φ) exceeding 0.74 stabilized by thin films of the aqueous phase, in which the high volume fraction of organic droplets are squeezed tightly together and deformed into polyhedral structures [1,2]. The particle-stabilized HIPEs, also known as Pickering HIPEs, have been widely used in the food, cosmetic, pharmaceutical, and petroleum industries nowadays [3,4]. From the perspective of the food industry, efforts to reduce the utilization of solid and semi-solid fats (e.g., butter and margarine) in past decades were prompted by the evidence that excessive consumption of saturated fatty acids and *trans*-fatty acids significantly increases the risk of cardiovascular disease [5], while protein-based HIPEs are promising replacements for them[6]. However, it has been a relatively broad consensus that the majority of proteins are "unqualified" HIPEs emulsifiers because of their poor amphiphilicity at the oil/water interface, on the basis of the theory of the Pickering particle framework [7].

Therefore, numerous efforts are devoted to increasing the amphiphilicity of proteins by surface modification, such as chemical cross-linking [4,8–11] and glycation [12], to achieve the goal of "**HIPEs stabilized solely by proteins**". Despite Xu and coworkers reporting that ovalbumin [13] and soy β-conglycinin [14] could directly stabilize HIPEs, they argued that the feasible mechanism is associated with the globular geometry and structural integrity of proteins [13]. Researchers now are facing a dilemma in defining the role of the proteins in the HIPEs formation. However, the overwhelming majority of researchers in colloids and interfaces science communities are usually concerned about mechanisms on the microscopic scale thus far [15], there were few experimental attempts to investigate the mesoscopic fluid dynamics in HIPEs because the emulsification in a rotor-stator mixer usually is considered as a turbulent flow regime [16].

In this work, we found that food-grade proteins, e.g., sodium caseinate (SC), whey protein isolation (WPI), and gelatin, could directly stabilize HIPEs by using the rotator-stator mixer. To our knowledge, these proteins were thought to be difficult to stabilize HIPEs [7], but here we uncovered an unusual pathway during the emulsification process for protein-based HIPEs stabilization related to the fluid dynamics. Furthermore, WPI, SC and gelatin are commercial food-grade materials, which are low-cost and abundantly available in practical, making it possible to manufacture HIPEs in industrial production-scale by using such a one-step method.

## 2 Materials and methods

### 2.1 Materials

Casein sodium salt from bovine milk (SC), gelatin from porcine skin (GL, type A, gel strength ~ 300 g bloom) and Nile Red were purchased from Sigma-Aldrich (St. Louis, MO, USA). Whey protein (WP, purity ≥ 80 %), β-lactoglobulin (purity ≥ 90%) and medium-chain triglycerides were purchased from Yuan-ye Biotech. Co., Ltd. (Shanghai, China). Sunflower oil was purchased from local market. All other chemi-cals used were analytical reagents.

## 2.2 Preparation of high internal phase emulsions

Protein solutions with different concentration (c = 0.2, 0.5, 0.75, 1.0, 2.0 and 3.0 % wt.) were prepared at ambient temperature, respectively, and stored overnight at 4 °C to enhance hydration. The emulsification process was prepared by adding the sunflower oil with varying fractions (Φ=0.70, 0.75, 0.80, 0.85 and 0.90) to the protein solution, then the mixture was homogenized by using a Fluko high-speed homogenizer (FA25, Shanghai, China) with a 25 mm head at 10000 r.p.m. for 1 min. In the experimental section, protein solution mixed with the sunflower oil.

## 2.3 Visualization techniques

### 2.3.1 High-speed camera

Visualization was performed by high-speed camera (pco. dimax HD, Germany) along with a 50 000 lm LED light source (Ryobi One+) as an illumination. The recording was done at 2128 fps with resolution 1920 X 1080 pixels. Shutter time was set to 1 ms.

### 2.3.2 Optical microscopy

The microscopic morphology of the Pickering HIPEs was observed using a Nikon optical microscope (Eclipse E100, Japan) with a 40× magnification glass.

### 2.3.3 Confocal laser scanning microscopy (CLSM)

Confocal laser scanning microscopy (CLSM) (Leica SP8, Carl Zeiss, Germany) was used to characterize the microstructure of the Pickering HIPEs. The Pickering

HIPEs were dyed with Nile Red (0.5 mg/mL), of which was excited by an argon lase with a wavelength of 552 nm. Microstructure of stained Pickering HIPEs was observed with an argon laser with a 63x magnifying glass. The image resolution was 1024 pixels.

## 3  Results and discussion

Red rectangular zones in Fig. 1a represent high internal phase emulsions (HIPEs) solely stabilized by whey protein isolation (WPI), sodium caseinate (SC) and gelatin, respectively. Its semi-solid texture is often evidenced by inverting the vial and observing a lack of flow under the force of gravity (Fig. 1b). Therefore, Fig. 1a illustrates stable HIPEs with internal phase volume ($\varphi$) ranging from 0.75 to 0.85, and both WPI and SC concentrations of 0.50-3.00 wt.%. Moreover, stable gelatin-based HIPEs were formed at $\varphi_{0.7}$ below the critical threshold of $\varphi_{0.74}$. In addition, such stable HIPEs was also formed in the case of β-lactoglobulin stabilizing medium-chain triglycerides (Supplementary materials, Figure S1).

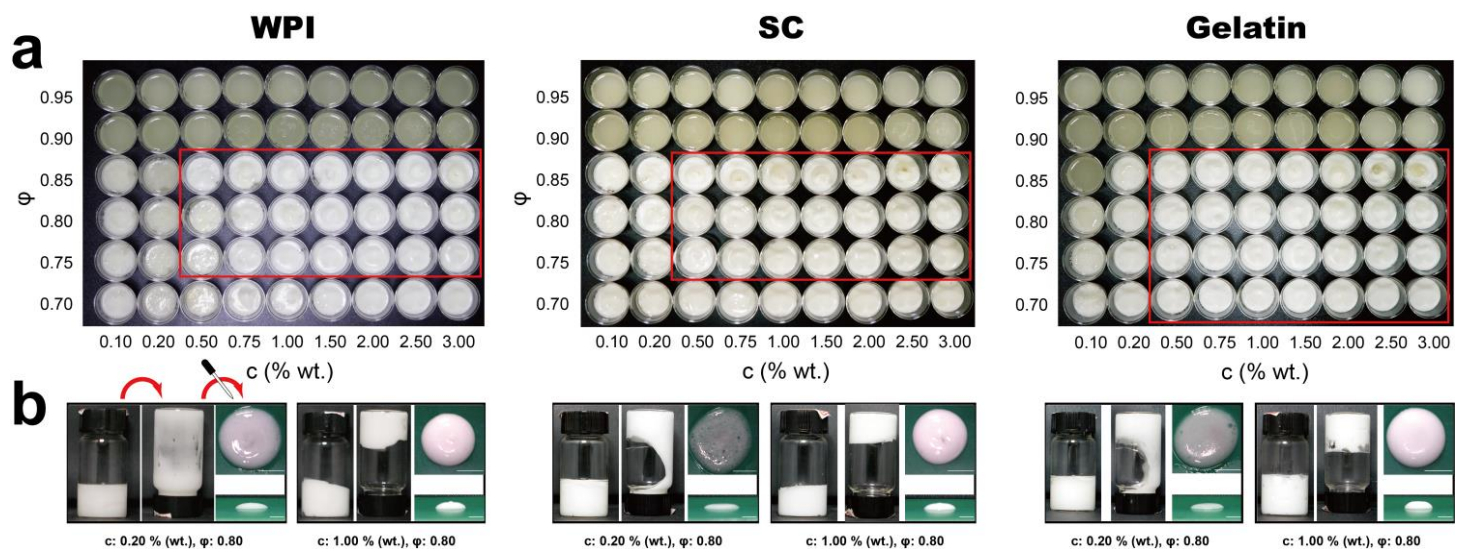

Fig. 1 (a) Visual appearance of high internal phase emulsions (HIPEs) stabilized by different concentrations of whey protein isolation (WPI), sodium caseinate (SC) and gelatin, respectively. (b) Photographs of HIPEs with volume oil fraction (Φ) 0.80 stabilized by WPI, SC and gelatin at the concentration of 0.20 and 1.00 wt.%.

The desired HIPEs were further validated by using confocal laser scanning microscopy (Fig. 2a). The oil droplets (stained by Nile Red) are squeezed into polygonal shapes. Moreover, the surface-averaged radius (**R**) of droplets decreased with protein concentrations (**c**) increasing, and exhibited relatively monodisperse in size (Supplementary materials, Figure S2). However, further **c** increment has little effect on the droplet size and no significant difference in **R** was observed among three protein-based HIPEs after 1.0 wt.% (Fig. 2b & Figure S2a). It demonstrates the effective concentration corresponding to the number of proteins covering the oil/water interface.

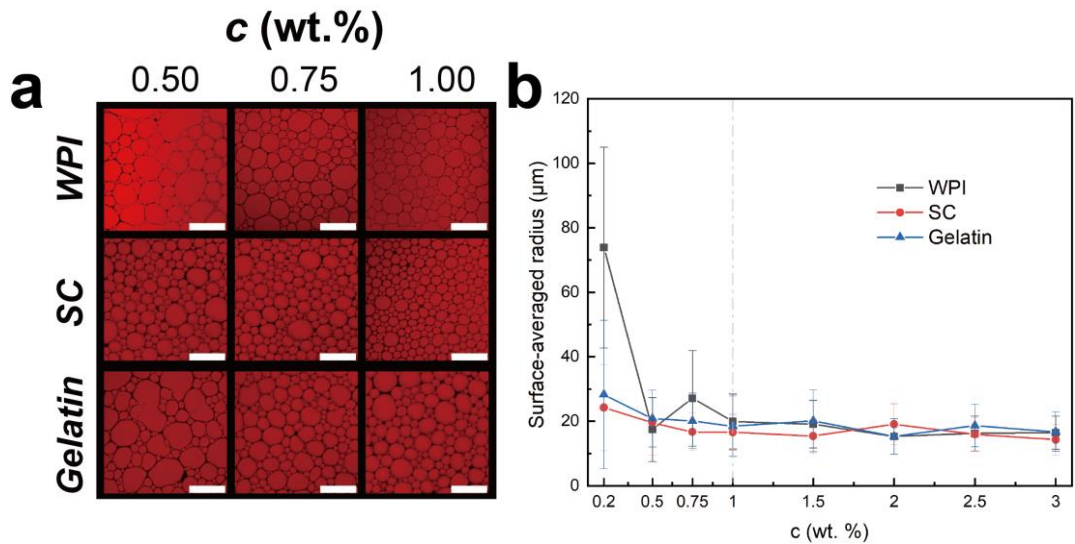

Fig. 2 Confocal laser scanning microscopy (CLSM) images of High internal phase emulsions (HIPEs) stabilized by different concentrations of (a) whey protein isolation (WPI), sodium caseinate (SC) and gelatin, respectively. HIPEs were stained by Nile red (color red). Scale bar: 50 μm. (b) The surface-average radius of droplet is stabilized by proteins (determined from the optical microscopy images and analyzed by Fiji-ImageJ software (version: 1.53c, National Institutes of Health, USA)).

From the mechanical processing view of a rotor-stator mixer device, the rotor rotates at high-speed inside the stator to generate high localized shear stress [16]. The main principle of the emulsification process is the continuous breaking of the droplets by turbulent eddies until a steady-state is reached [15]. Notably, the continuous flow behavior of the high internal phase ($\varphi_{0.8}$) emulsification process is significantly different from that of the medium internal phase ($\varphi_{0.5}$). In Fig. 3, we therefore proposed that the fluid transition is strongly influenced by $\varphi$.

Although similar emulsifying occurred in the case of HIPEs ($\varphi_{0.8}$) before t0+600 ms compared with that of the medium internal phase, after that, a peculiar transition from turbulent to laminar flow was observed. In Fig. 3b, a fluid laminarization around the mixer head was observed after t0+900 ms. As the stator and the beaker hamper the fluid motion, a downward tip streaming developed around the boundary of the cylinder, and an axisymmetric shape was presented (Fig. 3c).

It is well known that flow behaviors are dependent fundamentally on the Reynolds number (**Re**), which is a dimensionless number expressing the ratio of the inertial forces to the viscous forces. The phenomena of the fluid laminarization and axisymmetric vortices around the head is the evidence for the existence of a relatively low Re during the HIPEs emulsification process [17], which is dominated by viscous forces.

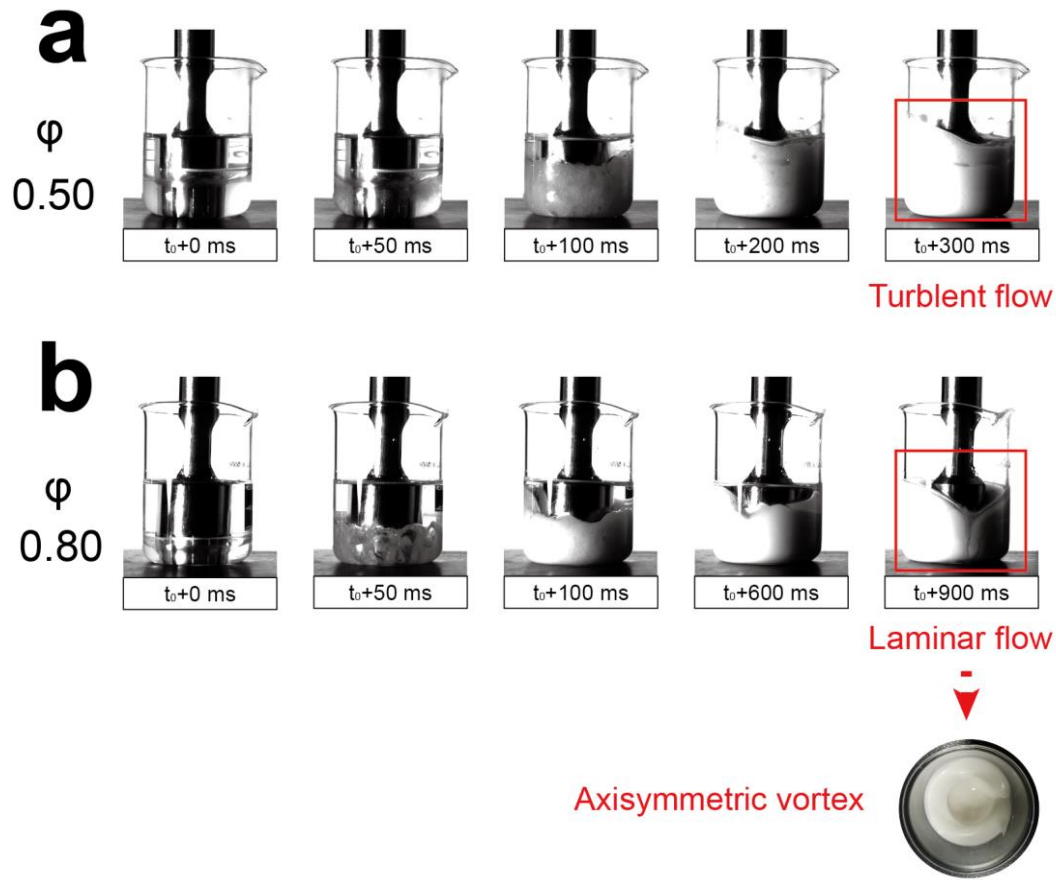

Fig. 3 High-speed photo of emulsification processing for the high internal phase ($\varphi_{0.8}$) and medium internal phase ($\varphi_{0.5}$) emulsions. (a) turbulent flow and (b) fluid laminarization were observed. (c) The axisymmetric vortex from the top view at the end of HIPEs processing.

We attempt to illustrate the spatiotemporal evolution of the HIPEs emulsification process, but it can be said that the mechanism is even more complex (Fig. 4). Since the two fluids are immiscible, the biphasic mixing caused by instantaneously fluid jets could be interpreted by the Richtmyer–Meshkov instability [18], which describes how a sudden intense acceleration set a lighter fluid in motion downward a heavier fluid through interface perturbation (Fig. 4a). Meanwhile, fluid jets also cause droplet breakup adjacent to the head, while the turbulent eddies derived from inertial forces predominantly dominate peripheral droplet breakup (Fig. 4b) [19].

Subsequently, although the interfacial tension gradients cause tangential stresses that produce fluid downward motions [20], this stress is rapidly attenuated by the action of dissipation through turbulent mixing. At the same time, a non-uniform velocity gradient in this process generated the laminar flow relating to the polyhedral droplets as a consequence of high momentum dissipation (Fig. 4c).

Since the creation and maintenance of organized non-equilibrium structures are due to dissipative processes in the HIPEs emulsification process, they are called dissipative structures and coined by Prigogine in 1969 [21]. The dissipative systems tend to coincide with an increased rate of entropy production required for the maintenance of such structures [22]. As the HIPEs emulsification in the rotor-stator mixer is a non-linear, thermodynamically open and non-equilibrium process, we found that the turbulent kinetic energy, externally high energy input provided *via* the rotor, dissipated irreversibly, which leads to the spontaneous organization, i.e., polyhedral droplets formation, and then the laminarization was observed (Fig. 4d). Similarly, the well-known paradigm of dissipative structure in fluid dynamics is the Rayleigh-Bénard (RB) convection, which demonstrated that the fluid is divided into horizontal cylindrical convection cells and rotates vertically in parallel plates driven by a thermal gradient [23].

Nevertheless, how estimate the threshold of self-organization's emergence from the non-equilibrium systems? The answer in the RB convection is the critical thermal difference which is determined by the dimensionless Rayleigh number [24]. In this scenario, we argue that $\varphi$ is the primary threshold. Since the theoretical calculation, $\varphi_{0.74}$, defines the maximum volume fraction of rigid uniform spheres filling a given volume efficiently [25], experimental evidence demonstrated that the close-packing for compressive droplets are polyhedral structures. The transition from sphere to polyhedral structure increases the contact surface area inevitably, at which HIPEs as a "jammed material" has a viscoelastic behavior with increasing friction forces (Fig. 4c) [26]. Hence, according to the dissipative structure theory, we propose that the viscous force emergency as a consequence of spontaneously organized structure dissipates the inertial force derived from turbulent kinetic energy, leading to the fluid laminarization.

In contrast to the dissipative structure in RB convection disappearing as soon as the heating stops, the dissipative structure in HIPEs increases the apparent viscosity, and thin aqueous films prevent droplet coalescence as well. On the other hand, droplets keep being squeezed and experience a constant Laplace pressure gradient in the steady-state (Fig. 4e) [27]. It has two balanced forces corresponding to the capillary number (**Ca**): the interfacial tension, which contributes towards keeping the drop's spherical form, and the dispersed phase viscosity is responsible for increasing the resistance of the internal drop medium to flow, delaying its deformation [28]. Hence, the presence of proteins alters drop shapes is a steady solution to the low **Re** drop breakup problem [29]. Since both water and oil are incompressible Newtonian fluids, proteins at the interface would form dense viscoelastic films to maintain the system stability and provide yield stress to mitigate the tension stress.

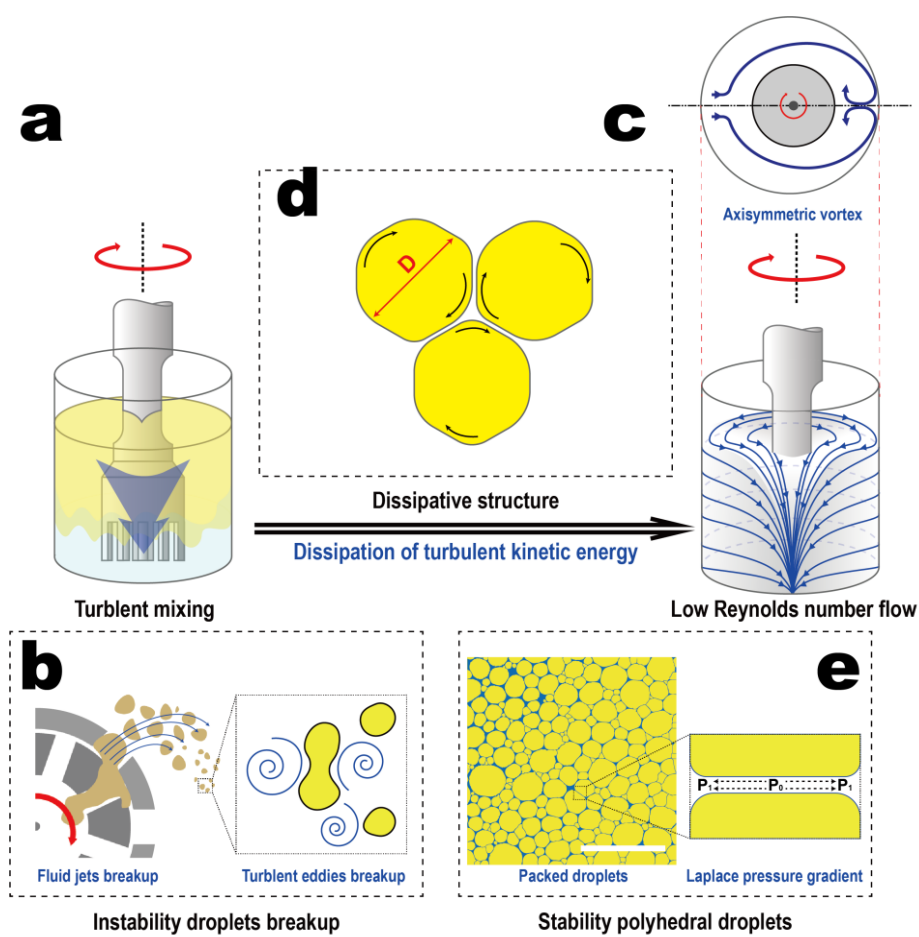

Fig. 4 Spatiotemporal development of protein-based HIPEs continuous process and stability. (a) Richtmyer–Meshkov (RM) instability interprets the initial downward motion. (b) Droplets breakup in the fluid jets and subsequently by turbulent eddies. (c) presents the streamline demonstrating the downward tip streaming laminar flow and axisymmetric vortex. (d) demonstrates the self-organized structures dissipate the turbulent kinetic energy. (e) demonstrates the crowded deformed droplets and the interfacial tension acting between the two fluid phas-

es. Pressure in the center ($P_0$) is higher than $P_1$, thus resulting in the fluid draining and this drainage flow applies to the stress to the surface, which can eventually result in coalescence [27].

## 4 Conclusion

In conclusion, this article uncovered a ubiquitous approach to stabilizing HIPEs directly by food-grade proteins, which are apparently different from the Pickering emulsions. As here some empirical results have been presented, the dissipative structure and the threshold φ are responsible for the appearance of fluid laminarization. Based on the dissipative structure theory, we argue that proteins can stabilize HIPEs through self-organization in such a non-equilibrium system, regardless of their amphiphilicity. Where an elastic protein film prevents the coalescence of the polyhedral droplets. It adds a new paradigm to presently existing dissipative structures and is expected to enlighten the industrial production-scale HIPEs production.

**Acknowledgments**

The research was supported by the National Natural Science Foundation of China (Grant No. 31871830) and the Natural Science Foundation of Zhejiang Province (Grant No. LY18C200002).

**References**

[1] V.O. Ikem, A. Menner, A. Bismarck, High Internal Phase Emulsions Stabilized Solely by Functionalized Silica Particles, Angew. Chemie Int. Ed. 47 (2008) 8277–8279. doi:10.1002/anie.200802244.

[2] N.R. Cameron, D.C. Sherrington, High internal phase emulsions

(HIPEs) — Structure, properties and use in polymer preparation, in: Biopolym. Liq. Cryst. Polym. Phase Emuls., Springer Berlin Heidelberg, Berlin, 1996: pp. 163–214. doi:10.1007/3-540-60484-7_4.

[3] I. Capron, B. Cathala, Surfactant-Free High Internal Phase Emulsions Stabilized by Cellulose Nanocrystals, Biomacromolecules. 14 (2013) 291–296. doi:10.1021/bm301871k.

[4] Z. Li, M. Xiao, J. Wang, T. Ngai, Pure Protein Scaffolds from Pickering High Internal Phase Emulsion Template, Macromol. Rapid Commun. 34 (2013) 169–174. doi:10.1002/marc.201200553.

[5] V.S. Shramko, Y. V. Polonskaya, E. V. Kashtanova, E.M. Stakhneva, Y.I. Ragino, The Short Overview on the Relevance of Fatty Acids for Human Cardiovascular Disorders, Biomolecules. 10 (2020) 1127. doi:10.3390/biom10081127.

[6] Y.-T. Xu, T. Yang, L.-L. Liu, C.-H. Tang, One-step fabrication of multifunctional high internal phase pickering emulsion gels solely stabilized by a softer globular protein nanoparticle: S-Ovalbumin, J. Colloid Interface Sci. 580 (2020) 515–527. doi:10.1016/j.jcis.2020.07.054.

[7] Q. Zhao, F. Zaaboul, Y. Liu, J. Li, Recent advances on protein-based Pickering high internal phase emulsions (Pickering HIPEs): Fabrication, characterization, and applications, Compr. Rev. Food Sci. Food Saf. 19 (2020) 1934–1968. doi:10.1111/1541-4337.12570.

[8]     X.-L. Li, W.-J. Liu, B.-C. Xu, B. Zhang, Simple method for fabrication of high internal phase emulsions solely using novel pea protein isolate nanoparticles: Stability of ionic strength and temperature, Food Chem. 370 (2022) 130899. doi:10.1016/j.foodchem.2021.130899.

[9]     B. Jiao, A. Shi, Q. Wang, B.P. Binks, High-Internal-Phase Pickering Emulsions Stabilized Solely by Peanut-Protein-Isolate Microgel Particles with Multiple Potential Applications, Angew. Chemie Int. Ed. 57 (2018) 9274–9278. doi:10.1002/anie.201801350.

[10]    S. Chen, L.-M. Zhang, Casein nanogels as effective stabilizers for Pickering high internal phase emulsions, Colloids Surfaces A Physicochem. Eng. Asp. 579 (2019) 123662. doi:10.1016/j.colsurfa.2019.123662.

[11]    H. Dai, Y. Li, L. Ma, Y. Yu, H. Zhu, H. Wang, T. Liu, X. Feng, M. Tang, W. Hu, Y. Zhang, Fabrication of cross-linked β-lactoglobulin nanoparticles as effective stabilizers for Pickering high internal phase emulsions, Food Hydrocoll. 109 (2020) 106151. doi:10.1016/j.foodhyd.2020.106151.

[12]    Y.-T. Xu, C.-H. Tang, B.P. Binks, High internal phase emulsions stabilized solely by a globular protein glycated to form soft particles, Food Hydrocoll. 98 (2020) 105254. doi:10.1016/j.foodhyd.2019.105254.

[13]    Y.-T. Xu, C.-H. Tang, T.-X. Liu, R. Liu, Ovalbumin as an Outstanding Pickering Nanostabilizer for High Internal Phase Emulsions, J.

Agric. Food Chem. 66 (2018) 8795–8804. doi:10.1021/acs.jafc.8b02183.

[14] Y.-T. Xu, T.-X. Liu, C.-H. Tang, Novel pickering high internal phase emulsion gels stabilized solely by soy β-conglycinin, Food Hydrocoll. 88 (2019) 21–30. doi:10.1016/j.foodhyd.2018.09.031.

[15] K.F. Kapiamba, Mini-review of the microscale phenomena during emulsification of highly concentrated emulsions, Colloid Interface Sci. Commun. 47 (2022) 100597. doi:10.1016/j.colcom.2022.100597.

[16] M. Gallassi, G.F.N. Gonçalves, T.C. Botti, M.J.B. Moura, J.N.E. Carneiro, M.S. Carvalho, Numerical and experimental evaluation of droplet breakage of O/W emulsions in rotor-stator mixers, Chem. Eng. Sci. 204 (2019) 270–286. doi:10.1016/j.ces.2019.04.011.

[17] J.H. Spurk, N. Aksel, Creeping Flows, in: Fluid Mech., Springer International Publishing, Cham, 2020: pp. 503–524. doi:10.1007/978-3-030-30259-7_13.

[18] Y. Zhou, R.J.R. Williams, P. Ramaprabhu, M. Groom, B. Thornber, A. Hillier, W. Mostert, B. Rollin, S. Balachandar, P.D. Powell, A. Mahalov, N. Attal, Rayleigh–Taylor and Richtmyer–Meshkov instabilities: A journey through scales, Phys. D Nonlinear Phenom. 423 (2021) 132838. doi:10.1016/j.physd.2020.132838.

[19] N. Vankova, S. Tcholakova, N.D. Denkov, V.D. Vulchev, T. Danner, Emulsification in turbulent flow, J. Colloid Interface Sci. 313 (2007) 612–629.

doi:10.1016/j.jcis.2007.04.064.

[20]   J. Chen, J. Wang, Z. Deng, X. Liu, Y. Chen, Experimental study on Rayleigh-Bénard-Marangoni convection characteristics in a droplet during mass transfer, Int. J. Heat Mass Transf. 172 (2021) 121214. doi:10.1016/j.ijheatmasstransfer.2021.121214.

[21]   D. Kondepudi, I. Prigogine, Modern Thermodynamics, Wiley, Weinheim, 2014. doi:10.1002/9781118698723.

[22]   D.K. Kondepudi, B. De Bari, J.A. Dixon, Dissipative Structures, Organisms and Evolution, Entropy. 22 (2020) 1305. doi:10.3390/e22111305.

[23]   A. Chatterjee, T. Ban, G. Iannacchione, Evidence of local equilibrium in a non-turbulent Rayleigh–Bénard convection at steady-state, Phys. A Stat. Mech. Its Appl. 593 (2022) 126985. doi:10.1016/j.physa.2022.126985.

[24]   F. Dabbagh, F.X. Trias, A. Gorobets, A. Oliva, Flow topology dynamics in a three-dimensional phase space for turbulent Rayleigh-Bénard convection, Phys. Rev. Fluids. 5 (2020) 024603. doi:10.1103/PhysRevFluids.5.024603.

[25]   K.. Lissant, The geometry of high-internal-phase-ratio emulsions, J. Colloid Interface Sci. 22 (1966) 462–468. doi:10.1016/0021-9797(66)90091-9.

[26]   R. Foudazi, S. Qavi, I. Masalova, A.Y. Malkin, Physical chemistry of highly concentrated emulsions, Adv. Colloid Interface Sci. 220 (2015) 78–

91. doi:10.1016/j.cis.2015.03.002.

[27] S. Moon, J.Q. Kim, B.Q. Kim, J. Chae, S.Q. Choi, Processable Composites with Extreme Material Capacities: Toward Designer High Internal Phase Emulsions and Foams, Chem. Mater. 32 (2020) 4838–4854. doi:10.1021/acs.chemmater.9b04952.

[28] F. De Vita, M.E. Rosti, S. Caserta, L. Brandt, On the effect of coalescence on the rheology of emulsions, J. Fluid Mech. 880 (2019) 969–991. doi:10.1017/jfm.2019.722.

[29] H.A. Stone, Dynamics of Drop Deformation and Breakup in Viscous Fluids, Annu. Rev. Fluid Mech. 26 (1994) 65–102. doi:10.1146/annurev.fl.26.010194.000433.

**Supplementary materials**

**Fluid laminarization in protein-based high internal phase emulsions process**

*Liang Guo[a]\*, Zi-an Deng[c], Yue-cheng Meng[b], Jing Chen[a], Sheng Fang,[b] Yang Pan[a] and Jie Chen[b]*

a   School of pharmacy, Nanjing University of Chinese medicine, 138 Xianlin Avenue, Nanjing, 210023, China

b   School of Food Science and Biotechnology, Zhejiang Gongshang University, 18 Xuezheng Road, Hangzhou, 310018, China

c   College of Agriculture & Biotechnology, Zhejiang University, 866 Yuhangtang Road, Hangzhou, 310058, China

\* Corresponding author

mingfangguo@njucm.edu.cn (L. Guo)

# β-lactoglobulin

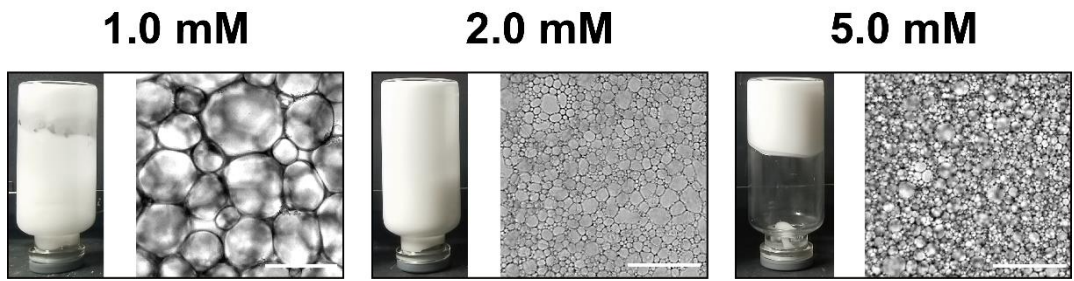

1.0 mM        2.0 mM        5.0 mM

Fig. S1 Photographs and microstructure of HIPEs with volume oil fraction (Φ) 0.80 stabilized by β-lactoglobulin at the concentration of 1.0, 2.0 and 5.0 mM respectively. Scale bar: 100 μm.

The supplemental image Figure S2a for Fig. 2b demonstrates that increasing the concentrations of protein from 1.00 to 3.00 wt.% does not lead to smaller droplets, suggesting that the size of the oil droplet is independent of the number of proteins when its concentration is large than 1.00 wt.%. Figure S2b and c illustrated that the droplet size does not decrease when the φ increased from 0.70 to 0.85. Nevertheless, the gelatin-based HIPEs is stiffer than that of WPI and SC at φ=0.70.

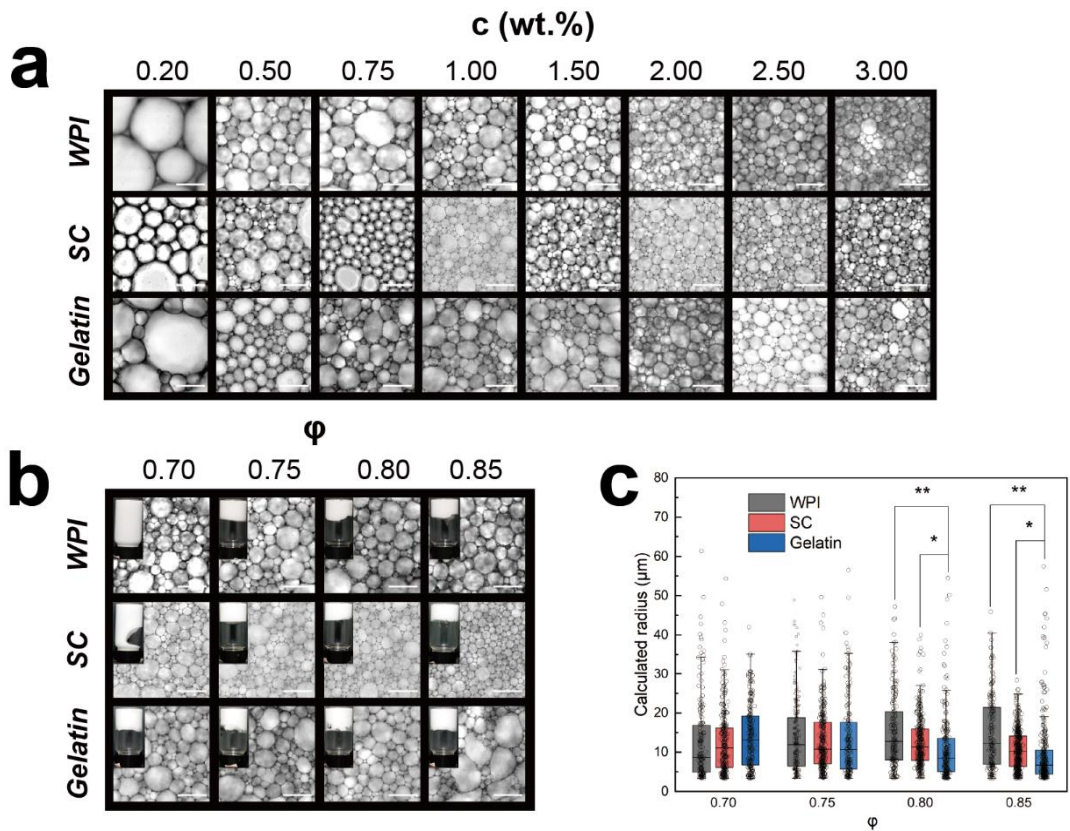

Fig. S2 (a) Microstructures of High internal phase emulsions (HIPEs) stabilized by different concentrations of whey protein isolation (WPI), sodium caseinate (SC) and gelatin (0.20, 0.50, 1.00 and 3.00 wt.%), respectively. (b) Microstructures and visual appearance of HIPEs with variant internal phase volume ($\Phi$) stabilized by WPI, SC and gelatin at a concentration of 1.00 wt.%. Scale bar: 100 μm. (c) Droplet size distribution of HIPEs with variant $\Phi$ stabilized by proteins at a concentration of 1.00 wt.%. (The average radius of droplets was determined from the optical microscopy images and analysed by Fiji-ImageJ software (version: 1.53c, National Institutes of Health, USA)).